\documentclass[epj]{svjour}
\usepackage{graphicx}
\usepackage{amsmath}
\usepackage{bm}
\usepackage{amssymb}
\usepackage{latexsym}
\usepackage{dcolumn}
\usepackage{url}

\newcolumntype{d}{D{.}{.}{10}}
\begin{document}
\title{Relativistic $K$ shell decay rates and fluorescence yields for Zn, Cd and Hg}
\author{C.\ Casteleiro\inst{1} \and
				F.\ Parente\inst{3} \and
        P.\ Indelicato\inst{2} \and
        J.\ P.\ Marques\inst{1}
%
}                     
\offprints{J.\ P.\ Marques}          
\institute{Centro de F{\'\i}sica At{\'o}mica e Departamento F{\'\i}sica, 
Faculdade de Ci{\^e}ncias, Universidade de Lisboa, \\
Campo Grande, Ed. C8, 1749-016 Lisboa, Portugal, 
\email{jmmarques@fc.ul.pt}
\and
Laboratoire Kastler Brossel,
\'Ecole Normale Sup\' erieure; CNRS; Universit\' e P. et M. Curie - Paris 6\\
Case 74; 4, place Jussieu, 75252 Paris CEDEX 05, France,
\email{paul.indelicato@spectro.jussieu.fr}
\and
Centro de F{\'\i}sica At{\'o}mica da Universidade de Lisboa e Departamento F{\'\i}sica da 
Faculdade de Ci{\^e}ncias e Tecnologia da Universidade Nova de Lisboa, Monte da Caparica, 2825-114 Caparica,  Portugal,
\email{facp@fct.unl.pt}
}
\date{Received: \today / Revised version: date}
%

\abstract{
In this work we use the multiconfiguration Dirac-Fock method to calculate the transition probabilities for all possible decay channels, radiative and radiationless, of a $K$ shell vacancy in Zn, Cd and Hg atoms. The obtained transition probabilities are then used to calculate the corresponding fluorescence yields which are compared to existing theoretical, semi-empirical and experimental results.
} 
\maketitle
%

\section{Introduction}
\label{Sec:Introd}
The fluorescence yield of an atomic shell or subshell is defined as the probability that a vacancy in that shell or subshell is filled through a radiative transition. An atom with a vacancy is in  an 
excited one-hole state. The fluorescence yield of an atomic shell or subshell is given by
\begin{equation}
        \omega_i=\frac{\Gamma_{R_i}}{\Gamma_{R_i}+\Gamma _{NR_i}}
\label{yield}
\end{equation}
where $\Gamma_{R_i}$ is the radiative width of an one-hole in that shell or subshell state and $\Gamma_{NR_i}$ is the radiationless width of the same state.

A good knowledge of K X-ray fluorescence yields is important for the interpretation of a large variety of measurements in fields such as nuclear, atomic, molecular and plasma physics as well as in medical physics, engineering and astrophysics \cite{handbook}.

Fluorescence yields have been object of intense research for several decades, both experimentally and theoretically, and a large number of articles have been published on this subject, describing different 
methods of measurement and calculation. In 1972, Bambynek \textit{et al.} \cite{Bambynek72} published a review article with a selection of the most reliable experimental $\omega_K$ values for 25 elements with $13\leq Z\leq 92$. In 1979, Krause  \cite{Krause79} published a compilation of $\omega_K$  values for 
all elements in the range $5 \leq Z \leq 110$, using for that both experimental and theoretical data. More recently (1994), Hubbell \textit{et al.} \cite{Hubbel94} compiled experimental and 
semi-empirical K shell fluorescence yields for elements with $11 \leq Z \leq 99$. Theoretical values of $\omega_K$ for elements with $4 \leq Z \leq 54$ were calculated with the Hartree-Fock-Slater 
method by McGuire in 1969 and 1970 \cite{McGuire69,McGuire70} and Walthers and Balla in 1971 \cite{Walthers71}. In 1980 Chen \textit{et al.} \cite{Chen80a} performed a relativistic Dirac-Hartree-Slater calculation of $\omega_K$ for selected elements with $18 \leq Z \leq 96$. Although  these data are already several decades old, they are still widely used for comparison with experimental results. 

In the last few years there has been a considerable increase in the number of experimental high precision measurements of the K shell fluorescence yields for many elements \cite{Ertugral07,Apaydin05,Yashoda05,Sahin05,Gunn03,Ozdemir02,Durak01,Simsek00,Simsek02}. The results show  several discrepancies when compared with theoretical and semi-empirical data. For Cd, for example, four experimental results have been published in this decade \cite{Yashoda05,Gunn03,Ozdemir02,Simsek02}. Yet, no relativistic calculation has been performed so far for this element. So, we decided to perform this calculation with state of art methods. We extended this calculation to Zn and Hg, which have electronic configurations similar to Cd. For Zn and Hg, the old relativistic results of Chen \cite{Chen80a}, using a simpler Dirac-Slater approach, can be used for comparison, and four experimental results have been published  in the Zn case \cite{Yashoda05,Gunn03,Durak01,Simsek00}, and one for Hg \cite{Apaydin05}, in the last eight years.

In this work we use the Multi-Configuration Dirac-Fock code of Desclaux and Indelicato~\cite{Desclaux75,mcdf}, which includes QED corrections, to calculate the K-shell fluorescence yields for the elements Zn, Cd and Hg. This is, to our knowledge, the first full relativistic calculation of this quantity for these elements. We note, that for Zn and Hg, Chen \textit{et al.} presented relativistic values, but they where obtained using a simpler relativistic Dirac-Slater approach~\cite{Chen80a}.

%
%
\section{Radiative and Radiationless  Transitions}

A general relativistic  program developed by Desclaux and Indelicato~\cite{mcdf}  was used to compute the energies and wavefunctions, as well as radiative and radiationless transition probabilities concerning the K fluorescence yields for Zn, Cd and Hg. Details on the Hamiltonian and the processes used to build the wavefunctions can be found elsewhere \cite{Desclaux75,Desclaux93,Indelicato95,Indelicato96}.  

For the radiative transition probabilities the code implements exact relativistic formulas  \cite{Gorceix87,Indelicato87,Indelicato98,Desclaux75,Indelicato89}. For the\-se transitions, the so-called Optimized Levels (OL) method was used to determine the  wavefunction and energy for each state involved. This method allows for a full relaxation of both initial and final states, providing much better energies and wavefunctions. However, spin-orbitals in the initial and final states are not orthogonal, since they have been optimized separately. To account for the wavefunctions nonorthogonality, the code uses the formalism  described by L\"{o}wdin~\cite{Lowdin55}.

In what concerns radiationless transitions, we assume that the creation of the inner-shell hole is independent of the decay process. In this way the primary ejected electron in the inner-shell ionization process does not interact with the Auger electron and radiationless transitions probabilities are calculated from perturbation theory. The continuum electron wavefunctions were obtained by solving the DF equations with the same atomic potential of the initial state. With this treatment, the continuum wavefunctions are made orthogonal to the initial bound state wavefunctions and the orthogonality of the wavefunctions is assured. No orbital relaxation is included in this calculation. The interaction potential between the core electron and the Auger electron include both Coulomb and magnetic parts. The Auger interaction two-electron operator is chosen to be the Breit operator~\cite{Santos99}.


\section{Results and Discussion}

In Table \ref{tab:rad_rates} we list the  radiative transition probabilities for all electron  transitions from higher subshells that could fill a vacancy in the K shell for Zn, Cd and Hg. In the first column  the final state vacancy subshell is identified. Our results for the total radiative transition probabilities agree closely with the ones obtained by Scofield \cite{Scofield75}, which are shown in the same table.

\begin{table}[ht]
\centering
\caption{K shell radiative transition probabilities for Zn, Cd and Hg (in a.u.). In the first column  we identify the final state subshell vacancy.}                                 
\begin{tabular}{lccc}                                                                  
&	Zn	&	Cd	&	Hg	\\
\hline		
\vspace{-3mm} \\
L1	&	2.59E-08	&	3.23E-06	&	7.852E-04	\\
L2	&	8.77E-03	&	6.47E-02	&	5.577E-01	\\
L3	&	1.71E-02	&	1.22E-01	&	9.462E-01	\\
M1	&	5.24E-09	&	8.73E-07	&	2.506E-04	\\
M2	&	1.19E-03	&	1.14E-02	&	1.077E-01	\\
M3	&	2.33E-03	&	2.21E-02	&	2.084E-01	\\
M4	&	1.59E-11	&	8.10E-05	&	2.766E-03	\\
M5	&	2.17E-06	&	1.13E-04	&	3.396E-03	\\
N1	&	3.50E-10	&	1.79E-07	&	7.002E-05	\\
N2	&		&	2.01E-03	&	2.623E-02	\\
N3	&		&	3.88E-03	&	5.165E-02	\\
N4	&		&	1.15E-05	&	7.557E-04	\\
N5	&		&	1.58E-05	&	9.235E-04	\\
N6	&		&		&	1.021E-06	\\
N7	&		&		&	1.152E-06	\\
O1	&		&	1.70E-08	&	1.440E-05	\\
O2	&		&		&	4.831E-03	\\
O3	&		&		&	9.090E-03	\\
O4	&		&		&	9.090E-03	\\
O5	&		&		&	1.035E-04	\\
O6	&		&		&	1.703E-06	\\
\hline
\vspace{-3mm} \\							
Total	&	2.94$\times 10^{-02}$	&	2.26$\times 10^{-01}$	&	1.93$\times 10^{+00}$	\\  
Scofield~\cite{Scofield75}	&	2.90$\times 10^{-02}$	&	2.26$\times 10^{-01}$	&	1.94$\times 10^{+00}$	\\        
\end{tabular}
\label{tab:rad_rates}
\end{table}

In Table \ref{tab:auger_rates} we present K shell radiationless transition probabilities from an initial state with a vacancy in the K shell to a final state with two vacancies for Zn, Cd and Hg. In the first column, for example, L1-LMN  means a final state with a first vacancy in the $L1$ subshell and the second vacancy in either  the L, M, or N shells. The disagreement between our results and the ones from Chen \textit{et al.} \cite{Chen80} for Zn and Hg, presented in the same table,  result, naturally, from the different methods employed, namely a full Dirac-Fock calculation and an approximate Dirac-Hartree-Slater calculation in Chen's work.

In Table \ref{tab:yields} we compare the fluorescence yields obtained in this work for  Zn, Cd and Hg 
with other theoretical and experimental data. We note that the present relativistic results are in general lower then existing non-relativistic theoretical data and are close to Chen \textit{et al.} \cite{Chen80a} relativistic values. In the same table we also list the semi-empirical values of Bambynek \textit{et al.} \cite{Bambynek72}, Krause  \cite{Krause79}, and Hubbell \textit{et al.} \cite{Hubbel94}. For Zn, these values lie systematically below most theoretical as well as experimental data.

\begin{figure*}
	\centering
		\includegraphics[width=12cm]{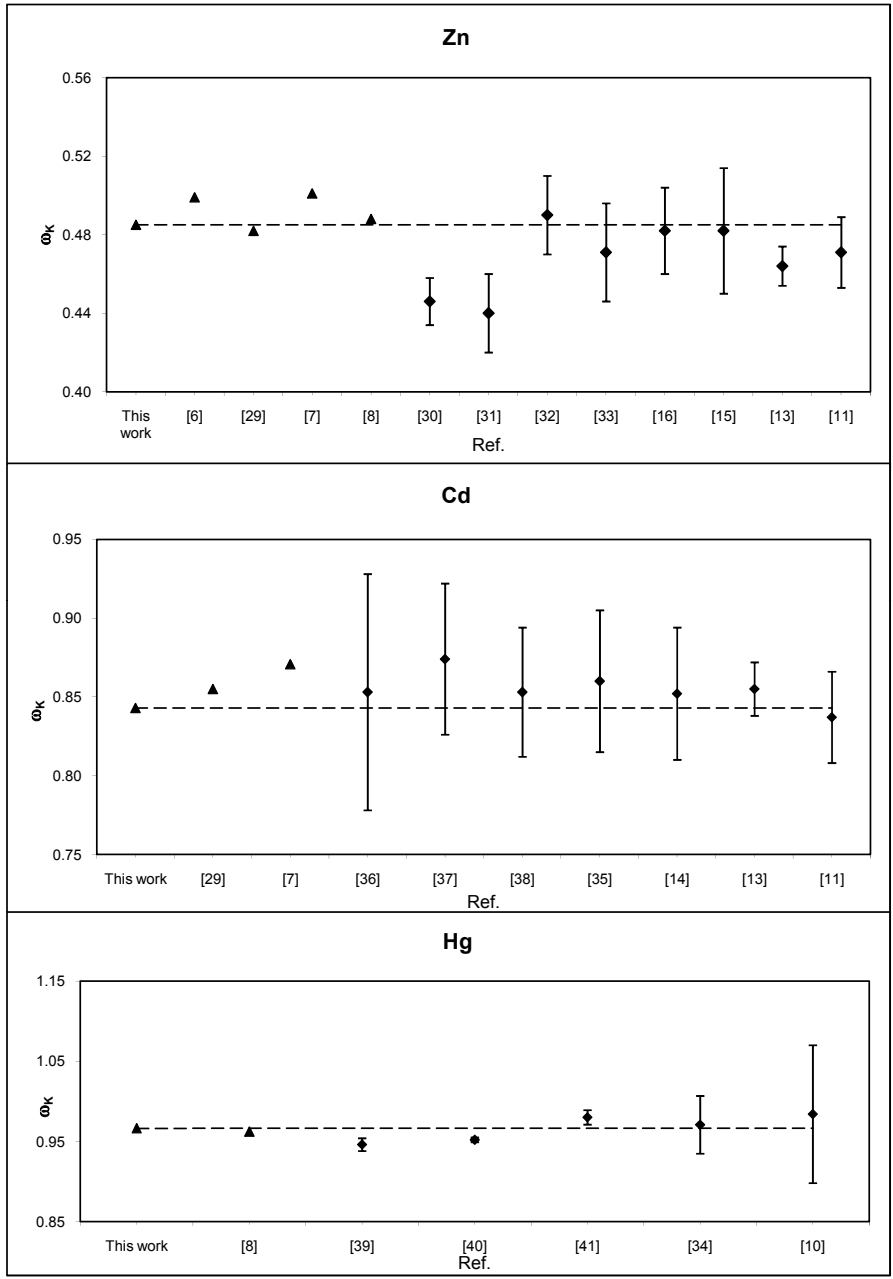}
	\caption{Comparison between experimental ($\blacktriangle$) and calculated  (${\blacklozenge}$)K shell fluorescence yields for Zn, Cd and Hg.}
	\label{figura}
\end{figure*}

In Figure \ref{figura} existing theoretical and experimental fluorescence yield values for Zn, Cd anf Hg are graphically compared with the results obtained in this work. In what concerns experimental data, our results agree, within the error bars, with the more recent experimental values except for the  Gudennavar \textit{et al.} Zn value. We note that the the pre-1960 experimental results for Hg present much smaller error bars than the more recent ones.

We can conclude that the precision of existing experimental data is not good enough to distinguish between the different theoretical values, thus calling for more precise experiments, in order to stimulate new calculations.


\section*{Acknowledgments}

This research was partially supported by the FCT research unit 303, financed by the European Community Fund FEDER, and by the French-Portuguese collaboration (PESSOA Program, Contract n$^{\circ}$ 10721NF). Laboratoire Kastler Brossel is Unit{\'e} Mixte de Recherche du CNRS,  de l'{\'E}cole Normale Sup{\'e}rieure et de l'Universit{\'e} Pierre et Marie Curie, n$^{\circ}$ C8552. Paul Indelicato acknowledges the support of the
Helmholtz Alliance Program of the Helmholtz Association, contract HA-216 ``Extremes of Density and Temperature: Cosmic Matter in the Laboratory''.



\begin{table*}
\centering
\caption{K shell radiationless transition probabilities for Zn, Cd and Hg (in a.u.). For example, L1-LMN stands for a final state with one vacancy in the L1 subshell and other vacancy in either of the L, M, and N shells. The initial state has one vacancy in the K shell.}        
\begin{tabular}{lclclc}
	&	Zn	&		&	Cd	&		&	Hg	\\
\hline
\vspace{-3mm} \\									
L1-LMN	&	1.10$\times 10^{-02}$	&	L1-LMNO	&	1.63$\times 10^{-02}$	&	L1-LMNOP	&	3.74$\times 10^{-02}$	\\
L2-LMN	&	1.53$\times 10^{-02}$	&	L2-LMNO	&	1.53$\times 10^{-02}$	&	L2-LMNOP	&	1.78$\times 10^{-02}$	\\
L3-LMN	&	4.78$\times 10^{-03}$	&	L3-LMNO	&	4.15$\times 10^{-02}$	&	L3-LMNOP	&	8.83$\times 10^{-03}$	\\
M1-MN	&	1.47$\times 10^{-04}$	&	M1-MNO	&	4.34$\times 10^{-04}$	&	M1-MNOP	&	1.50$\times 10^{-03}$	\\
M2-MN	&	2.06$\times 10^{-04}$	&	M2-MNO	&	3.87$\times 10^{-04}$	&	M2-MNOP	&	7.30$\times 10^{-04}$	\\
M3-MN	&	2.57$\times 10^{-05}$	&	M3-MNO	&	2.43$\times 10^{-04}$	&	M3-MNOP	&	6.08$\times 10^{-04}$	\\
M4-MN	&	8.91$\times 10^{-07}$	&	M4-MNO	&	2.60$\times 10^{-04}$	&	M4-MNOP	&	3.34$\times 10^{-05}$	\\
M5-MN	&	2.15$\times 10^{-07}$	&	M5-MNO	&	7.01$\times 10^{-06}$	&	M5-MNOP	&	2.93$\times 10^{-05}$	\\
N1-N	&	1.48$\times 10^{-07}$	&	N1-NO	&	1.09$\times 10^{-05}$	&	N1-NOP	&	7.82$\times 10^{-05}$	\\
	&		&	N2-NO	&	2.23$\times 10^{-06}$	&	N2-NOP	&	3.47$\times 10^{-05}$	\\
	&		&	N3-NO	&	1.34$\times 10^{-06}$	&	N3-NOP	&	2.60$\times 10^{-05}$	\\
	&		&	N4-NO	&	8.28$\times 10^{-08}$	&	N4-NOP	&	1.85$\times 10^{-06}$	\\
	&		&	N5-NO	&	2.12$\times 10^{-08}$	&	N5-NOP	&	2.85$\times 10^{-05}$	\\
	&		&	O1-O	&	2.37$\times 10^{-08}$	&	N6-NOP	&	1.12$\times 10^{-08}$	\\
	&		&		&		&	N7-NOP	&	2.23$\times 10^{-08}$	\\
	&		&		&		&	O1-OP	&	2.29$\times 10^{-06}$	\\
	&		&		&		&	O2-OP	&	7.91$\times 10^{-07}$	\\
	&		&		&		&	O3-OP	&	3.91$\times 10^{-07}$	\\
	&		&		&		&	O4-OP	&	8.22$\times 10^{-09}$	\\
	&		&		&		&	O5-OP	&	2.83$\times 10^{-09}$	\\
	&		&		&		&	P1-P	&	9.01$\times 10^{-09}$	\\
\hline
\vspace{-3mm} \\											
Total	&	3.14$\times 10^{-02}$	&		&	7.44$\times 10^{-02}$	&		&	6.71$\times 10^{-02}$	\\
Chen \textit{et al.} \cite{Chen80}	&	3.04$\times 10^{-02}$	&		&		&		&	7.48$\times 10^{-02}$	\\
\end{tabular}
\label{tab:auger_rates}
\end{table*}

\begin{table*}
\centering
\caption{K shell flourescence yields for Zn, Cd and Hg.}
\protect\label{tab:yields}                    
               
\begin{tabular}{llll}                      												
	&	Zn			&	Cd			&	Hg			\\
\\											
	&				&	Theoretical							\\
\hline													
This work	&	0.485			&	0.842			&	0.966			\\
McGuire \cite{McGuire70}	&	0.499			&				&				\\
Kostroun \cite{Kostroun71}	&	0.482			&	0.855			&				\\
Walters \textit{et al.} \cite{Walthers71}	&	0.501			&	0.871			&				\\
Chen \cite{Chen80a}	&	0.488			&				&	0.962			\\
\\											
	&				&	Semi empirical							\\
\hline													
Hubbel \cite{Hubbel94}	&	0.469			&	0.843			&	0.966			\\
Bambynek \cite{Bambynek72}	&	0.479			&	0.841			&	0.965			\\
Krause \cite{Krause79}	&	0.474			&	0.836			&	0.980			\\
\\													
	&				&					Experimental							\\
\hline	
Broyles \textit{et al.}\cite{Broyles53}	&				&				&	0.946	$\pm$	0.008	\\
Roos \cite{Roos57}	&	0.446	$\pm$	0.012	&				&				\\
Patronis \textit{et al.} \cite{Patronis57}	&	0.44	$\pm$	0.02	&				&				\\
Nall \textit{et al.} \cite{Nall60}	&				&				&	0.952	$\pm$	0.003	\\
Arora \textit{et al.} \cite{Arora81}	&	0.49	$\pm$	0.02	&				&				\\
Bhan \textit{et al.} \cite{Bhan81}	&				&	0.853	$\pm$	0.075	&				\\
Al-Nasr \textit{et al.} \cite{Al-Nasr87}	&				&	0.874	$\pm$	0.048	&				\\
Sidhu  \textit{et al.} \cite{Sidhu88}	&				&				&	0.980	$\pm$	0.009	\\
Piuos \textit{et al.} \cite{Pious92}	&	0.471	$\pm$	0.025	&				&				\\
Balakrishna \textit{et al.} \cite{Balakrishna94}	&				&	0.853	$\pm$	0.041	&				\\
Durak \textit{et al.} \cite{Durak98}	&				&				&	0.971	$\pm$	 0.036	\\
Sim\c{s}ek \textit{et al.} \cite{Simsek00}	&	0.482	$\pm$	0.022	&				&				\\
Durak \textit{et al.} \cite{Durak01}	&	0.482	$\pm$	0.032	&				&				\\
Sim\c{s}ek \textit{et al.}\cite{Simsek02}	&				&	0.860	$\pm$	0.045	&				\\
\"{O}zdemir \textit{et al.} \cite{Ozdemir02}	&				&	0.852	$\pm$	0.042	&				\\
Gudennavar \textit{et al.}\cite{Gunn03}	&	0.464	$\pm$	0.010	&	0.855	$\pm$	0.017	&				\\
Yashoda \textit{et al.} \cite{Yashoda05}	&	0.471	$\pm$	0.018	&	0.837	$\pm$	0.029	&				\\
Apaydin \textit{et al.} \cite{Apaydin05}	&				&				&	0.984	$\pm$	0.086	\\
\end{tabular}\end{table*}

\newpage

\end{document}